\documentstyle [prl,aps]{revtex}

\def\sech{\,\hbox{sech}}

\begin{document}
\title{
  \begin{flushright} \begin{small}
    hep-th/0104101
  \end{small} \end{flushright}
\vspace{1.cm}
Stability Analysis of Anisotropic Inflationary Cosmology}
\author{Chiang-Mei Chen\footnote{E-mail: cmchen@phys.ntu.edu.tw}
and W. F. Kao\footnote{E-mail: wfgore@cc.nctu.edu.tw}}
\address{$^*$Department of Physics, National Taiwan University,
Taipei 106, Taiwan \\ and \\
$^\dagger$Institute of Physics, Chiao Tung University, Hsinchu,
Taiwan}
\date{July 20, 2001}
\maketitle

\begin{abstract}
The stability analysis of an anisotropic inflationary 
universe of the four dimensional Neveu-Schwarz--Neveu-Schwarz
string model with a nonvanishing cosmological constant is discussed in this paper. The accelerating
expansion solution found earlier is shown to be stable against
the perturbations with respect to the dilaton and axion fields
once the dilaton field falls close to the local minimum of the
symmetry-breaking potential. This indicates that the Bianchi I
space tends to evolve to an isotropic flat
Friedmann-Robertson-Walker space. This expanding solution is also
shown to be stable against the perturbation with respect to
anisotropic spatial directions.
\end{abstract}


\section{Introduction}
Observations of the cosmic microwave background radiation,
galaxies and other astronomical objects reveals that our
universe, on a very large scale, is remarkably uniform
\cite{data,cobe,barrow} and is currently under accelerated
expansion \cite{nova,nova1,cluster,cluster1}. Therefore, a
physical acceptable cosmology should provide a self-contained
mechanism to smear out the primordial anisotropy and achieve an
accelerating expansion at present time. Furthermore, it is firmly
believed that the Einstein's theory of general relativity breaks
down at high enough energy which, at least, occurred during the
early epoch of our universe. It is known that the string theories
are promising candidates for all particles and interactions,
including the gravity, in a unified formulation. Thus the
astrophysical and cosmological implications of string theories
have become a highly developing research themes (see \cite{LWC00}
and references therein).

An anisotropic cosmology was considered recently in the framework
of four-dimensional Neveu-Schwarz--Neveu-Schwarz (NS-NS) effective
string theory in which the gravitational field is coupled with the
dilaton and axion fields \cite{CHM01}. In the de Sitter geometry
configuration, i.e. with a positive cosmological constant
$\Lambda>0$, an inflationary solution can be found in an exact
parametric form. At large time limit, this universe is
isotropilized and it's expansion is accelerating which is
consistent with our current astronomical observations. Stability
analysis of any solution may provide more information of the
theory under consideration
\cite{jb1,jb2,kp91,dm95,dm2,KPZ99,kao00,kim,abel}. Therefore it is
interesting, and is the purpose of the present paper, to
investigate the stability conditions of this anisotropic
inflationary string cosmology. Our analysis indicates that the
inflationary solution found in Ref. \cite{CHM01} remains {\em
stable} when the potential $U(\phi)$ is close to the local minima
of the potential.

This paper is organized as follows. In Sec. II we briefly review
the NS-NS effective string theory and the exact solution found in
\cite{CHM01}. The stability analysis with respect to the
perturbations of the dilaton and axion fields will be performed
in Sec. III. The perturbation with respect to the gravitational
field will be studied in Sec. IV. In Sec. V, we will draw some
conclusions.

\section{Anisotropic Solution}
The four-dimensional NS-NS effective action, which is common to
both the heterotic and the type II string theories \cite{LWC00},
is given by
\begin{equation} \label{SE}
{\cal S} = \int d^4x \sqrt{-g} \left\{ R - \kappa (\partial
\phi)^2 - \frac1{12} e^{-4\phi} \, H_{[3]}^2 - U \right\},
\end{equation}
where $\phi$ is the so-called dilaton and $H_{\mu\nu\lambda} =
\partial_{[\mu}B_{\nu\lambda]}$ is a totally anti-symmetric
tensor. Moreover, the $\kappa$ denotes the generalized dilaton
coupling constant ($\kappa=2$ for typical superstring theories)
and $U=U(\phi)$ is a dilaton potential. The field equations of
the action (\ref{SE}) can be shown to be
\begin{eqnarray}
R_{\mu\nu} - \kappa \partial_\mu \phi \partial_\nu \phi - \frac12
g_{\mu\nu} U - \frac14 e^{-4\phi}\left( H_{\mu\alpha\beta}
H_\nu{}^{\alpha\beta} - \frac13 g_{\mu\nu} H^2 \right) &=& 0,
\label{EqR} \\
\nabla_\mu \left( e^{-4\phi} H^{\mu\nu\lambda} \right) &=& 0,
\label{EqH} \\
\nabla^2 \phi + \frac1{6\kappa} e^{-4\phi} H^2 - \frac1{2\kappa}
\partial_\phi U &=& 0. \label{Ephi}
\end{eqnarray}
In addition, the totally anti-symmetric tensor $H$ obeys the
Bianchi identity $\partial_{[\mu} H_{\nu\lambda\rho]}=0$.

In four dimensions, any three-form can be mapped one-to-one onto
its dual one-form. This mapping relates the three rank tensor $H$
to a pseudo-vector $A$ via $H={}^* A$ with $A=A_\mu dx^\mu$ the
dual one-form. Moreover, one can show that the tensor $H$ can be
solved to give
\begin{equation}\label{Deh}
H^{\mu\nu\lambda} =  \, e^{4\phi} \,
\epsilon^{\mu\nu\lambda\rho} \, \partial_\rho h
\end{equation}
following the field equation (\ref{EqH}). Here the totally
anti-symmetric tensor $\epsilon$ denotes the Levi-Civita tensor
and $h=h(t)$ is known as the Kalb-Ramond axion. Note that we use
the convention that
$\epsilon^{\mu\nu\lambda\rho}=-\delta^\mu_{[0} \delta^\nu_1
\delta^\lambda_2 \delta^\rho_{3]}/\sqrt{-g}$ in this paper. In
addition, the above solution holds if and only if the first
homology group for the space considered is trivial
\cite{topo,topo1,topo2}. Consequently, the Bianchi identity shown
earlier becomes
\begin{equation}\label{Eh}
\partial_{\mu}\left(\sqrt{-g}\, e^{4\phi}\, \partial^\mu h\right)=0.
\end{equation}

We will consider the case that the potential of the dilaton field
is a positive constant
\begin{equation}\label{AU}
U(\phi) = \Lambda \ge 0,
\end{equation}
which represents a de Sitter space. An exact solution for an
anisotropic (Bianchi type I) cosmology has been found in
Reference \cite{CHM01}. Note that the line element of a Bianchi
type I space, an anisotropic generalization of the flat
Friedmann-Robertson-Walker (FRW) geometry, can be written as
\begin{equation}
ds^2 = -dt^2 + a_1^2(t) dx^2 + a_2^2(t) dy^2 + a_3^2(t) dz^2,
\end{equation}
with $a_i(t), i=1,2,3$ the expansion factors on each spatial
directions. With the identities (\ref{Deh}, \ref{AU}), the field
equations (\ref{EqR}, \ref{Ephi}, \ref{Eh}) will take the
following form
\begin{eqnarray}
3 \dot \theta + \sum_{i=1}^3 \theta_i^2 + \kappa \dot \phi^2 +
\frac12 e^{4\phi} \, \dot h^2 - \frac12 \Lambda &=& 0,
\label{dth} \\
\frac1{V} \partial_t ( V \theta_i ) - \frac12 \Lambda &=& 0,
\quad i=1,2,3, \label{dV} \\
\ddot h + 3 \theta \dot h + 4 \dot \phi \dot h &=& 0, \label{dh} \\
\frac1{V} \partial_t ( V \dot\phi ) - \frac1{\kappa} e^{4\phi}
\dot h^2 &=& 0 \label{dphi}.
\end{eqnarray}
Here we have introduced the {\em volume scale factor}, $V :=
\prod_{i=1}^3 a_i$, {\em directional Hubble factors}, $\theta_i :=
\dot a_i/a_i,\, i=1,2,3$, and the {\em mean Hubble factor},
$\theta := \sum_{i=1}^3 \theta_i/3 = \dot V/3V$ for convenience.
In addition, we will also introduce two basic physical
observational quantities in cosmology: the {\em mean anisotropy
parameter}, $A := \sum_{i=1}^3 (\theta_i-\theta)^2/3\theta^2$,
and the {\em deceleration parameter}, $q := \partial_t
\theta^{-1}-1$. Note that $A \equiv 0$ for an isotropic expansion.
Moreover, the sign of the deceleration parameter indicates how
the Universe expands. Indeed, a positive sign corresponds to
``standard'' decelerating models whereas a negative sign
indicates an accelerating expansion.

It was shown that the general exact solution of this model can be
parameterized in the following form \cite{CHM01}
\begin{eqnarray}
a_i(\tau) &=& a_{i0} \sinh^{\alpha_i^+} \frac{\tau}2
\cosh^{\alpha_i^-} \frac{\tau}2, \quad i=1,2,3, \\
e^{2\phi(\tau)} &=& \varphi_0^2 \left( \tanh^\omega\frac{\tau}2 +
\tanh^{-\omega}\frac{\tau}2 \right), \\
h(\tau) &=& h_0 + \frac{\kappa\sqrt{\phi_0}}{C} \,
\frac{\tanh^{2\omega}\frac{\tau}2-1}
{\tanh^{2\omega}\frac{\tau}2+1}.
\end{eqnarray}
Therefore, one has
\begin{eqnarray}
\theta(\tau) &=& \sqrt{\frac{\Lambda}6} \coth \tau, \\
V(\tau) &=& V_0 \sinh \tau,
\end{eqnarray}
where $\tau := \sqrt{3\Lambda/2}(t-t_0),\, \alpha_i^\pm := 1/3
\pm \sqrt{2/3\Lambda}K_i/V_0,\, \omega :=
\sqrt{8\phi_0/3\Lambda}/V_0$ and
$\varphi_0^2:=\sqrt{C^2/8\kappa\phi_0}$. Here $t_0,\, a_{i0},
K_i,\, C \ge 0,\, \phi_0>0$ and $h_0$ are free parameters
representing the constants of integration. In addition, one
writes $V_0=\Pi_{i=1}^3 a_{i0}/2$ for convenience. One can also
show that there is an additional constraint $\sum_{i=1}^3 K_i=0$
following the field equations. These integration constants also
obey the consistency condition,
\begin{equation}
K^2 := \sum_{i=1}^3 K_i^2 = \Lambda V_0^2 - \kappa \phi_0,
\end{equation}
which follows from equation (\ref{dth}). Thereby the mean
anisotropy and the deceleration parameter are given by
\begin{eqnarray}
A(\tau) &=& \frac{2K^2}{\Lambda V_0^2} \sech^2 \tau, \\
q(\tau) &=& 3 \sech^2 \tau - 1.
\end{eqnarray}

This exact solution indicates that the evolution of the Bianchi
type I universe starts from a singular state, but with finite
values of the mean anisotropy and deceleration parameters. In the
large time limit the mean anisotropy tends to zero, $A \to 0$, and
the universe approaches an isotropic inflationary de Sitter phase
with a negative deceleration parameter, $q<0$, providing an
accelerating expanding universe consistent with the present
observations. Furthermore, in the large time limit the dilaton and
axion fields become constants, $\lim_{t\to\infty}h(t) = h_0 =
\text{constant}$, $\lim_{t\to\infty}e^{2\phi(t)} = 2\varphi_0^2 =
\text{constant}$. Note that the dynamics and evolution of the
universe is determined only by the presence of a cosmological
constant (or a dilaton field potential). There is no direct
coupling between the metric and the dilaton and axion fields other
than the constraint on the integration constants.

Note that the mixmaster model \cite{m1,m2,m3,m4,m5} discussed
earlier deals with a model with perfect fluid. Earlier work
focuses on the behavior of the solution near the singularity. The
static field equation studied in Ref. \cite{CHM01} is similar to
the mixmaster model with axion and dilaton behaving like the
perfect fluid. The solution shown in this section is, however, the
first time such exact solution is found for the NS-NS string
effective theory that exhibits the desired property driving an
initially anisotropic Bianchi type I space to an isotropic space.
We will focus on the stability property of this solution in the
following section.

\section{Perturbation and Stability}
Perturbations of the fields of a gravitational system against the
background evolutionary solution should be checked to ensure the
stability of the exact or approximated background solution. In
principle, the stability analysis should be performed against the
perturbations of all possible fields in all possible manners
subject to the field equations and boundary conditions of the
system. In the following section, we will divide the perturbations
into two disjoint classes: (a) the perturbations of the scale
factors, or equivalently the metric field, and the axion field;
and (b) the perturbations of the dilaton and axion fields.

Note further that the axion field can be solved as a combination
of the dilaton field and the scale factors according to Eq.
(\ref{dh}). The result is
\begin{equation} \label{ax}
\dot h = C \, e^{-4\phi} \, V^{-1},
\end{equation}
with a constant of integration $C \ge 0$. Here $C=0$ means that
the axion field does not couple to any other field. Therefore, the
effect of the perturbation of the axion field can be replaced by
the equation (\ref{ax}). In fact, Eq. (\ref{Edh}) indicates that
if one perturbs the axion field, one has to perturb either the
dilaton field or $a_i$ altogether unless the constant $C$
vanishes.  Note also that equation (\ref{ax}) indicates that $\dot
h \sim a^{-3}$ which is much smaller than $a$ in the large time
limit where $a_i \to a$ for all directions. This is an indication
that the effect of the axion field perturbation can be ignored in
the metric perturbation we will discuss later.

Once we replace the effect of the axion field in the field
equations (\ref{dth}), (\ref{dV}), and (\ref{dphi}), one can
further simplified the perturbations into two disjoint class: (a)
the perturbations of the scale factors and  (b) the perturbations
of the dilaton fields. We will argue that the most complete
stability conditions we are looking for can be obtained from class
(a) and class (b) perturbations even the backreaction of the
scalar field perturbation on the metric field perturbations is
known to be important \cite{back}. We will show that this
backreaction does not bring in any further restriction on the
stability conditions.

The reason is rather straightforward. One can write the linearized
perturbation equation as
\begin{equation}
D^{i}_{a_{j}} \delta a_j +D^i_\phi \delta \phi =0
\end{equation}
for the system we are interested. Here the axion perturbations
have been replaced according to Eq. (\ref{ax}). Moreover,
perturbations are defined as $a_i = a_i^0 + \delta a_i $ and $\phi
=\phi_0 + \delta \phi$ with the index $0$ denoting the background
field solution. Note also that the operators $D^i_{a_j}$ and
$D^i_{\phi}$ denote the differential operator one obtained from
the linearized perturbation equation with all fields evaluated at
the background solutions. The exact form of these differential
operators will be shown later in the following arguments.

One is looking for stability conditions that the field parameters
must obey in order to keep the evolutionary solution stable. One
can show that class (a) and class (b) solutions are good enough to
cover all domain of stability conditions. Let us denote the domain
of solutions to class (a), (b) and (a+b) stability conditions as
$S(a), S(b)$, and $S(a+b)$ respectively. Specifically, the
definition of these domains are defined by $S(a) \equiv \{ \delta
a_i | D^{i}_{a_{j}} \delta a_j =0 \}$, $S(b) \equiv \{ \delta \phi
| D^{i}_{\phi} \delta \phi =0 \}$, and $S(a+b) \equiv \{ ( \delta
a_i, \delta \phi ) | D^{i}_{a_{j}} \delta a_j + D^{i}_{\phi}
\delta \phi =0 \}$.

Therefore, one only needs to show that $S(a) \cap S(b) \subset
S(a+b)$. This is because that ``$D^{i}_{a_{j}} \delta a_j =0$ and
$D^{i}_\phi \delta \phi =0$'' imply that ``$D^{i}_{a_{j}} \delta
a_j +D^i_\phi \delta \phi =0$''. On the other hand,
``$D^{i}_{a_{j}} \delta a_j +D^i_\phi \delta \phi h=0$'' does not
imply that ``$D^{i}_{a_{j}} \delta a_j =0$ or $D^{i}_\phi \delta
\phi =0$''. Hence class (a) and class (b) solutions cover all the
required stability conditions we are looking for. Hence we only
need to consider these two separate cases for simplicity.

In addition, one knows that any small time-dependent perturbation
against the metric field is known to be equivalent to a gauge
choice \cite{gauge}. This can be clarified as follows. Indeed, one
can show that any small coordinate change of the form $x'^\mu =
x^\mu-\epsilon^\mu$ will induce a gauge transformation on the
metric field according to $g'_{\mu \nu}=g_{\mu \nu}+D_\mu
\epsilon_\nu +D_\nu \epsilon_\mu$. Therefore, a small metric
perturbation against a background metric is amount to a gauge
transformation of the form $a_i'=a_i+ \epsilon^t \dot a_i$ for the
Bianchi type I metric with $\epsilon_\mu =({\rm constant},
\epsilon_i(t))$. This is then equivalent to small metric
perturbations. If a background solution is stable against small
perturbation with respect to small field perturbations, one in
fact did nothing but a field redefinition.

If the background solution is, however, unstable against small
perturbations, e.g. the small perturbation will grow exponentially
as we will show momentarily, the resulting large perturbations can
not be classified as small gauge transformation any more.
Therefore, the stability analysis performed in the literatures
\cite{jb1,jb2,kp91,dm95,dm2,KPZ99,kao00,kim,abel} for various
models against unstable background solution served as a very
simple method to check if the system support a stable metric field
background. This is the reason why we still perform a perturbation
on the metric field for stability analysis even small perturbation
is equivalent to a gauge redefinition.


Note that one should also consider a more general perturbation
with space perturbation included. The formulation is, however,
much more complicate than the one we will show in this paper. We
will focus on the time-dependent case for simplicity in this
paper. The space-dependent perturbation analysis is still under
investigation. The time-dependent analysis alone will, however,
brings us many useful information for the stability conditions
about the model we are interested. For example, we will show in
the following subsection that the solution found in Ref.
\cite{CHM01} remains stable as long as the scalar field falls
close to any local minimum of the potential $U(\phi)$. Note again
that the solution found in Ref. \cite{CHM01} is an exact solution
only when $U=$ constant.

\subsection{Dilaton Field Perturbation}
In this subsection, we will consider the perturbation $\delta\phi$
with respect to the exact background solution $\phi_B$ given in
the previous section.

We will consider a general symmetry-breaking potential $U(\phi)$
which has at least one local minimum. The solution obtained in
section II remains a good approximation as long as the dilation
field is close to the local minimum ($\phi=v$) of the potential
$U$ such that $U_0:=U(v)=\Lambda$. Therefore, a perturbation on
$\phi$ field can also be considered as a test to see whether such
solution tends to stabilize the FRW space or not. We will show
momentarily that the result indicates that FRW space tends to be a
stable final state of any Bianchi type I space once the dilaton
field approaches the local minimum of the symmetry-breaking
potential.

Due to the fluctuation $\delta\phi$, the deviation of the dilaton
potential, up to the first order, is
\begin{equation}
U(\phi_B+\delta\phi) \to U(\phi_B) + \partial_\phi U(\phi_B) \,
\delta\phi = \Lambda  + \partial_\phi U(\phi_B) \, \delta\phi.
\end{equation}
By keeping the metric fields unperturbed, one can show that the
perturbation equations are
\begin{eqnarray}
\dot \phi_B \, \delta\dot\phi - \frac{C^2}{\kappa V_B^2}
e^{-4\phi_B} \, \delta\phi &=& 0, \\
\partial_\phi U(\phi_B) \, \delta\phi &=& 0, \label{pU}\\
\frac1{V_B} \partial_t (V_B \, \delta\dot\phi) + 4 \dot\phi_B \,
\delta\dot\phi + \frac1{2\kappa} \partial_\phi^2 U(\phi_B) \,
\delta\phi &=& 0.
\label{Edh}
\end{eqnarray}
Here we use the the Eq. (\ref{ax}) in above equations. In
addition, once the dilaton fields falls close enough to one of the
local minimum of the symmetry-breaking potential, the dynamics of
the dilaton perturbation is inevitably much smaller than the scale
factor perturbations. This confirm our claims earlier that class
(a) perturbation and class (b) perturbation are not of the same
order of magnitude.

Note also that the Eq. (\ref{pU}) indicates that one of the
necessary stable conditions for the background solution is that
$\phi_B$ should be a local minimum of the dilaton potential
$U(\phi)$. In fact, close to local minimum condition is the only
situation we are interested in this paper.

Combining the equations of $\delta\dot\phi$, one finds that the
dilaton perturbation $\delta\phi$ is determined by
\begin{equation} \label{dphione}
\partial_t \left( V_B e^{4\phi_B} \, \delta\dot\phi \right) +
\frac1{2\kappa} V_B e^{4\phi_B} \partial_\phi^2 U(\phi_B) \,
\delta\phi = 0.
\end{equation}
Taking the large time limit, $t\to\infty$, one can show that the
background variables $V_B$ and $\phi_B$ approach
\begin{equation}
V_B \to \frac{V_0}2 e^{\alpha t}, \qquad \phi_B \to v,
\end{equation}
where $\alpha := \sqrt{3\Lambda/2}$ and $v:=\ln(2\varphi_0^2)/2$.
In addition, the equation (\ref{dphione}) reduces to
\begin{equation}
\delta\ddot\phi + \alpha\,\delta\dot\phi + \beta\,\delta\phi = 0,
\end{equation}
with $\beta=\partial_\phi^2 U(\phi)|_{\phi=v}/2\kappa$. For the
case $\partial_\phi^2 U(\phi)|_{\phi=v}=0$, i.e. $\beta=0$, the
perturbations of dilaton and axion fields behave as $\delta\phi
\propto \exp(-\alpha t)$ and $\delta h \propto \exp(-2\alpha t)$
indicating both fields approach zero exponentially in the large
time limit. On the other hand, the dilaton perturbation can be
shown to be $\delta\phi \propto \exp(\gamma t)$ with
$\gamma=\left(-\alpha \pm \sqrt{\alpha^2-4\beta}\right)/2$ for
the case $\beta\ne 0$. This implies that $|\delta\phi|$ vanishes
rapidly near the local minimum $v$ when $\partial_\phi^2
U(\phi)|_{\phi=v}>0$. As well, $\delta h \propto
\exp[(\gamma-\alpha)t]$ also converge to zero exponentially.

In short, the necessary condition for the stability of the
background solutions is that the background dilaton field must be
close to the local minimum  $v$ of the dilaton potential
$U(\phi)$. Consequently, the perturbed fields will approach zero
exponentially provided that $\partial^2_\phi U(\phi)|_{\phi=v}$
is non negative. This partially reflects the fact that the
background solution remains a good approximated solution to the
system as long as the dilaton potential $U(\phi) \sim$ constant
near $\phi=v$.

Note that we know the evolutionary solution of the system only
when the dilaton potential is a constant. Nonetheless, the
argument shown in this subsection indicates that the system tends
to bring the universe to the FRW space once the dilaton field
falls close to local minimum of the symmetry-breaking potential as
$t \to \infty$. This result provides a convincing evidence for the
formation of the flat FRW space evolved from a Bianchi type I
anisotropic space-time.

\subsection{Metric Perturbation}
We will study the stability of the background solution with
respect to perturbations of the metric in this subsection.
Perturbations will be considered for all three expansion factors
$a_i$ via
\begin{equation}
a_i \to a_{Bi} + \delta a_i = a_{Bi} (1 + \delta b_i),
\end{equation}
hereafter. We will focus on the variables $\delta b_i$ instead of
$\delta a_i$ from now on for convenience. Therefore, the
perturbations of the following quantities can be shown to be
\begin{equation}
\theta_i \to \theta_{Bi} + \delta\dot b_i, \quad \theta \to
\theta_B + \frac13 \sum_i \delta\dot b_i, \quad \sum_i \theta_i^2
\to \sum_i \theta_{Bi}^2 + 2 \sum_i \theta_{Bi}\, \delta\dot b_i,
\quad V \to V_B + V_B \sum_i \delta b_i.
\end{equation}
One can show that the metric perturbations $\delta b_i$, to the
linear order in $\delta b_i$, obey the following equations
\begin{eqnarray} \label{h31}
\sum_i\delta\ddot b_i + 2\sum_i\theta_{Bi} \delta\dot b_i -
\frac{C^2}{ V_B^2}
e^{-4\phi_B} \, \sum_i \delta b_i &=& 0,\\
\delta\ddot b_i + \frac{\dot V_B}{V_B}\delta\dot b_i +
\sum_j\delta\dot b_j \, \theta_{Bi} &=& 0,\\
\sum_i\delta\dot b_i \, \dot\phi_B &=& 0. \label{h3}
\end{eqnarray}
Equation (\ref{h3}) indicates that $\sum_i \delta \dot b_i =0$.
Therefore, Eq. (\ref{h31}) gives further that
\begin{equation} \label{bi-con}
\sum_i \delta b_i = 0.
\end{equation}
Substituting it back to the other equations, one can show that
\begin{eqnarray}
\delta \ddot b_i + \frac{\dot V_B}{V_B}\delta\dot b_i &=& 0.
\end{eqnarray}

Note that the background variables $V_B$ and $\theta_{Bi}$
approach
\begin{equation}
V_B \to \frac{V_0}2 e^{\alpha t}, \qquad \theta_{Bi} \to
\sqrt{\Lambda/6},
\end{equation}
in the large time limit as $t \to \infty$. Here $\alpha :=
\sqrt{3\Lambda/2}$. Note also that we keep only leading order in
the above equations. Therefore, the solution for $\delta b_i$ can
be found to be
\begin{equation}
\delta b_i = c_i e^{-\alpha t},
\end{equation}
together with the following constraint on integration constants
$c_i$
\begin{equation}\label{CC}
\sum_i c_i = 0.
\end{equation}
Moreover, the asymptotic form, in the large time limit, of
background scale factors is $a_{Bi} \to a_{i0} \exp [ \alpha t /3
] /4^{1/3} $. Therefore, the ``actual'' fluctuations for each
expansion factor, $\delta a_i = a_{Bi} \delta b_i$ is
\begin{equation}
\delta a_i \to \frac{a_{i0} c_i}{4^{1/3}} e^{-2\alpha t/3}.
\end{equation}
Hence $\delta a_i$ approaches zero exponentially since $\alpha$
is definite positive. Consequently, the background solution is
{\em stable} against the perturbation of the graviton field.

Note that the perturbations of the three different expansion
factors in Bianchi type I cosmology are not independent due to
the constraint (\ref{CC}). The constraint (\ref{CC}) indicates
that scale factor perturbations along arbitrary directions are
confined by the field equations. This also signifies the symmetry
of the coordinate transformation among the scale factors. In
fact, there should be only two independent coordinate
perturbations reflecting the symmetry of the coordinate choice.

\section{Conclusion}
In summary, we investigate the stability condition for an
anisotropic inflationary string solution which evolves to the FRW
space. This solution is consistent with the significant
requirements constrained by present astronomical observations ---
homogeneity, isotropy and accelerating expansion. We analyze in
details the perturbed equations with respect to the dilaton fields
and separately with respect to the expanding scale factors of the
Bianchi type I space. Our result indicates that the cosmological
model considered in this paper is stable for both cases under
certain constraints.

In the former case, the necessary condition for the stability of
the background solutions is that the background dilaton field must
be close to the local minimum  $v$ of the dilaton potential
$U(\phi)$. Consequently, the perturbed fields will approach zero
exponentially provided that $\partial^2_\phi U(\phi)|_{\phi=v}$ is
non negative. This partially reflects the fact that the background
solution remains a good approximation to the system only when the
dilaton potential $U(\phi) \sim$ constant near $\phi=v$.

Note that we do not known the exact evolutionary solution of the
system when the dilaton potential is not close to a constant.
Nonetheless, we show that the system tends to bring the universe
to the FRW space once the dilaton field falls close to local
minimum of the symmetry-breaking potential as $t \to \infty$. This
result provides a convincing evidence for the formation of the
flat FRW space evolved from a Bianchi type I anisotropic
space-time which will work for models with a reasoable
symmetry-breaking potential.

For the later case, the perturbed expanding scale factors are
also shown to approach zero exponentially in the large time limit.
We also show that the perturbations of the three expanding scale
factors are constrained by the field equations.

\section*{Acknowledgments}
The work of CMC is supported by the Taiwan CosPA project and, in
part, by the Center of Theoretical Physics at NTU and National
Center for Theoretical Science. This work is supported in part by
the National Science Council under the grant number
NSC89-2112-M009-001. One of the authors (WFK) is grateful to the
hospitality of the physics department of National Taiwan
University where part of this work is done.

\end{document}